\documentclass[prl,aps,epsf,amsfonts,floats,twocolumn,amssymb,amsmath,groupedaddress,floatfix,nofootinbib]{revtex4}

\usepackage{graphicx}
\usepackage{subeqnarray}
\usepackage[]{latexsym}
\usepackage{bm}

\newcommand{\be}{\begin{equation}}\newcommand{\ee}{\end{equation}}
\newcommand{\bea}{\begin{eqnarray}}\newcommand{\eea}{\end{eqnarray}}
\newcommand{\bsa}{\begin{subeqnarray}}
\newcommand{\esa}{\end{subeqnarray}}
\newcommand{\brr}{\begin{array}}\newcommand{\err}{\end{array}}
\newcommand{\bit}{\begin{itemize}}\newcommand{\eit}{\end{itemize}}
\newcommand{\ben}{\begin{enumerate}}\newcommand{\een}{\end{enumerate}}

\newcommand{\ba}{\begin{array}}
\newcommand{\ea}{\end{array}}

\def\lab{\label}
\def\lf{\left}

\def\non{\nonumber}
\def\rar{\rightarrow}
\def\ri{\right}
\def\al{\alpha}\def\ga{\gamma}\def\Ga{\Gamma}

\def\ka{\kappa}

\def\Om{\Omega}

\def\1{{_{1}}}\def\2{{_{2}}}

\def\noHe0{:\;\!\!\;\!\!:H_e(0):\;\!\!\;\!\!:}
\def\noHm0{:\;\!\!\;\!\!:H_\mu(0):\;\!\!\;\!\!:}

\def\lab{\label}

\def\lf{\left}

\def\non{\nonumber}

\def\rar{\rightarrow}
\def\ri{\right}

\def\al{\alpha}\def\ga{\gamma}
\def\Ga{\Gamma}

\def\ka{\kappa}

\def\Om{\Omega}

\def\1{{_{1}}}\def\2{{_{2}}}

\begin{document}

\title{Thermal condensate structure and   cosmological energy density of the Universe  }

\author{ Antonio Capolupo}
\author{ Gaetano Lambiase}
\author{ Giuseppe Vitiello}
 \affiliation{ Dipartimento di Fisica "E.R.Caianiello",
  Universit\'a di Salerno, and INFN Gruppo collegato di Salerno, Fisciano (SA) - 84084, Italy}

\begin{abstract}

The aim of this paper is the study of thermal vacuum condensate for scalar and fermion fields. We analyze the   thermal states at the temperature of the cosmic microwave background (CMB) and we show that the vacuum expectation value of the energy momentum tensor density of photon fields reproduces the energy density and pressure of the CMB. We perform the computations in the formal framework of
 the thermo field dynamics. We also consider the case of neutrinos and thermal states at the temperature of the neutrino cosmic background.
Consistency with the estimated lower bound of the sum of the active neutrino masses is verified.    In the boson sector, non trivial contribution to the energy of the universe is given by  particles of masses of the order of $10^{-4}eV$  compatible with the ones of the axion-like particles.
 The fractal self-similar structure of the thermal radiation is also discussed and related to the coherent structure of the thermal vacuum.

\end{abstract}

\maketitle

\section{Introduction}

The task of this paper is the analysis of  the   thermal vacuum condensate for scalar and fermion fields, with specific reference to temperatures characteristic of cosmic microwave background (CMB).
The interest in considering the vacuum condensate in relation with  CMB  resides in the fact that  it is   a thermal radiation filling almost uniformly the observable universe and one expects that it plays a relevant role in the universe thermal vacuum stracture.
CMB appears as a radiation left over from an early stage in the expansion  of the universe~\cite{ref1} and
  has a thermal black body spectrum corresponding to the temperature of $2.72548 \pm 0.00057$ K \cite{ref2}.
The anisotropies contained in the spatial variation in the spectral density are attributed to small thermal variations, presumably generated by quantum fluctuations of matter \cite{ref1}, \cite{ref3}, \cite{ref4}.

In our analysis, we compute the expectation value of the energy momentum tensor density of photon fields on the thermal vacuum. As a result, we obtain the energy density and pressure of the CMB.

Together with the CMB, there is an indirect evidence of the existence of the cosmic neutrino background (CNB) which represents the universe's background particle radiation composed of neutrinos (relic neutrinos) \cite{Kolb-Turner}-\cite{Kolb-Turner3}. The CNB estimated temperature is roughly $1.95 K$~\cite{Kolb-Turner}. It is therefore interesting to extend our study of thermal vacuum condensate also to the CNB case.
Thus, we assume the hierarchical neutrino model and, by computing  the energy density of the neutrino thermal vacuum, we check the  lower bound of the sum of the active neutrino masses  $\sum m_{\nu}$, which has been  estimated from the  neutrinos  oscillations to be of the order of $0.06  eV$ \cite{Gonzales}.

%A lower limit of order of    it is set

% play a crucial role in the understanding of different phenomena of high energy physics and astrophysics \cite{Capolupo:2006et}- \cite{Capolupo:2010ek}. The discovery of neutrino oscillations has shown that neutrinos have mass.
 %Moreover, they make it possible to determine or limit all three masses from measurements of electron-flavor neutrinos in beta decay. The present upper limit from such measurements is $2 eV$.

%We will consider thermal a similar computation for Majorana fields and consider the temperature of the neutrino cosmic background.

 %   By assuming that the energy density of the neutrino thermal vacuum is bounded from above by the energy density of the photon thermal vacuum, we
% derive an upper bound on the absolute mass of the lightest neutrino, i.e. $m_{\nu,1} \lesssim 10^{-4}eV$.

%Moreover, by assuming  the hierarchical neutrino model and using the results of the experiments on solar and atmospheric neutrinos, from which the differences between the squared masses  $\Delta m_{12}^{2} $ and  $\Delta m_{23}^{2}$ have been obtained,  we derive the absolute masses of the other neutrinos and the sum of the active n\documentclass[10pt]{•}eutrino masses. We find a value of  $\sum m_{\nu}$ of order of $0.07 eV$, which is in agreement with the estimated lower bound on the neutrino masses.

% result which does not depend
%on whether the neutrino is a Dirac or Majorana particle.

We finally discuss the fractal self-similar structure of the thermal vacuum.

In Section II,  the Thermo Field Dynamics (TFD) formalism is introduced and the general expressions  of its energy density and pressure are shown.
Explicit computations for Maxwell, scalar and  fermion fields are presented in Section III and,
in Section IV, the fractal  structure of the thermal states is analyzed.
Section V is devoted to the conclusions.

\section{Thermal vacuum and particle condensate }

The thermal vacuum state $|0(\theta)\rangle$, with $\theta = \theta(\beta)$, $\beta \equiv 1/(k_{B}T)$ and $k_{B}$ the Boltzmann constant, is introduced in the  TFD  formalism~\cite{Takahashi:1974zn,Umezawa} in such a way that
 the thermal statistical average ${\cal N}_{a_{\bf k}}(\theta)$ is given by
  ${\cal N}_{a_{\bf k}}(\theta) = \langle 0(\theta)| N_{a_{\bf k}} |0(\theta)\rangle$, with $N_{a_{\bf k}} = a^{\dag}_{\bf k} a_{\bf k} $, the number operator. The bosonic operators $a_{\bf k}$ and $a^{\dag}_{\bf k}$   have usual canonical commutation relations (CCR).

  The explicit form of   $|0(\theta)\rangle$ is
\bea
\lab{(2.12)} |0(\theta)\rangle = \prod_{\bf k}
{1\over{\cosh{\theta_k}}} \exp{ \left ( \tanh {\theta_k} ~a_{\bf
k}^{\dagger} {b}_{\bf k}^{\dagger} \right )} |0\rangle \,,
\eea
and it is recognized  to be  a two-mode time dependent generalized $SU(1,1)$ coherent state~\cite{Perelomov:1986tf,Klauder}, condensate of pairs of  $a_{\bf k}$ and $b_{\bf k}$ quanta. $|0\rangle$ is  the vacuum annihilated by $a_{\bf k}$ and $b_{\bf k}$.
The auxiliary boson operator $b_{\bf k}$  commutes with $a_{\bf k}$  and is introduced in order to produce the trace operation in   computing thermal averages.  The thermal vacuum $| 0(\theta)  \rangle$  is normalized to one,  $ \langle 0(\theta) | 0(\theta) \rangle  = 1, ~ \forall~ \theta$~ and  in the infinite volume
limit $
{ \langle 0(\theta (\beta)) | 0 \rangle  \rightarrow 0~~ {\rm
as}~~ V\rightarrow \infty }, ~~~\forall~  \beta\,
$  (for $\int \! d^{3} \kappa ~\theta_{\kappa} $ finite
and~positive).

One also has  $ { \langle 0(\theta(\beta)) | 0(\theta(\beta'))  \rangle  \rightarrow 0~
{\rm as}~ V\rightarrow \infty},~ \forall~\beta$ and $\beta',  \beta' \neq \beta$. Thus  $\{ |0(\theta (\beta)) \rangle  \}$ provides a representation  of
the CCR  defined at each  {\it $\beta$} and  unitarily inequivalent $\forall~ \beta'\neq \beta$ to any other representation $\{ |0(\beta') \rangle \}$ in the infinite volume limit.

% This implies that at a given temperature $T$, in thermal equilibrium ($\beta$ constant in time $t$), the system sits in the representation of the CCR corresponding to such a $T$. In thermal non-equilibrium conditions, with temperature changing in time, i.e. $\beta = \beta(t)$, the system
%evolves in time through unitarily inequivalent representations of the CCR and it is known that such a time evolution is controlled by the entropy operator~\cite{Takahashi:1974zn,Umezawa}.

 Note that
$a_{\bf k}$ and $b_{\bf k}$ do not annihilate the state $| 0(\theta)  \rangle$. The annihilation operators, say $A_{\bf k}(\theta_k)$ and $B_{\bf k}(\theta_k)$,  for   $| 0 (\theta)  \rangle$, $A_{\bf k}(\theta_k)| 0(\theta) \rangle = 0 = B_{\bf k}(\theta_k)| 0(\theta) \rangle $,  are obtained through the Bogoliubov transformation
\bea\non
A_{\bf k}(\theta_k) &=& e^{i{\theta_k}{\cal G}}\, a_{\bf k}\,
e^{-i{\theta_k}{\cal G}} =  a_{\bf k} \,{\rm cosh} ~\theta_k - {b}^{\dagger}_{\bf k}
\, {\rm sinh} ~\theta_k,
\\\non
{B}_{\bf k}(\theta_k) &=& e^{i{\theta_k}{\cal G}} \,{b}_{\bf k}\,
e^{-i{\theta_k}{\cal G}}  = {b}_{\bf k} \,{\rm cosh} ~\theta_k -
 a^{\dagger}_{\bf k} \,{\rm sinh} ~\theta_k\,,
\lab{A7}
\eea
whose generator  $\cal G$ is given by ${\cal G} =- i \sum_{\bf k}{ ( a_{\bf
k}^{\dagger} {b}_{\bf k}^{\dagger} - a_{\bf k} {b}_{\bf
k}) }$.
The thermal vacuum expectation value of the number operator $N_{a_{\bf k}}= a_{\bf k}^{\dag}a_{\bf k}$ is given by
\bea \lab{(2.13a)}
\hspace{-4mm}{\cal N}_{a_{\bf k}}(\theta) &=& \langle 0(\theta)
|a_{\bf k}^{\dagger}a_{\bf k}| 0(\theta)\rangle =  \sinh^{2} \theta_k.
 %\\ \non
% &=&\frac{1}{e^{\beta(t) E_k}-1}\,.
 \eea

Minimization of the free energy (see below) then leads to the thermal statistical average of $N_{a_{\bf k}}$
\bea \lab{(2.13a)}
\hspace{-4mm}{\cal N}^{B}_{a_{\bf k}}(\theta) = \sinh^{2} \theta_k = \frac{1}{e^{\beta {\Omega}_k}-1}\,,
 \eea
which is  indeed the Bose-Einstein distribution function for $a_{\bf k}$.
%Similarly, for fermions one obtains $| 0(\theta)\rangle =
%\prod_{\bf k} (\cos {\theta_k} + \sin {\theta_k} ~a_{\bf
%k}^{\dagger} {b}_{\bf k}^{\dagger})| 0\rangle$, where $a_{\bf
%k}$ and $ {b}_{\bf k}$ are anticommuting fermion operators, and
%%{1\over{\cosh{\theta_k}}} \exp{ \left ( \tanh {\theta_k} ~a_{\bf
%%k}^{\dagger} {b}_{\bf k}^{\dagger} \right )} |0\rangle .$ the state
%%
%\bea \lab{(2.13b)}
%\hspace{-4mm}{\cal N}^{F}_{a_{\bf k}}(\theta) = \sin^{2} \theta_k = \frac{1}{e^{\beta {\Omega}_k} + 1}\,.
% \eea
% %

Summing up,  the ``thermal background'' at   $T$ is described by the quantum coherent condensate vacuum  $| 0(\theta)\rangle$, which is the thermal physical vacuum.

We now are ready to compute  the contributions of the energy momentum tensor $T^{\mu \nu}$ to the thermal vacuum  for Maxwell, scalar and  fermion fields.
 We observe that the off-diagonal terms of $T^{\mu \nu}$ on the vacuum state are zero  for these fields, i.e. $\langle 0(\theta )|T^{i j}(x)|0(\theta )\rangle = 0$, for $i \neq j$.
Therefore, the vacuum condensate is homogenous and isotropic and behaves as a perfect fluid (similar result hold for mixed particles \cite{Capolupo:2006et}-\cite{Capolupo:2006et5} and for curved space \cite{maroto}).
Then  the  energy density and pressure induced by the  condensate (\ref{(2.13a)}), at a given time (we consider the red shift $z$ of the universe), can be defined  by computing the  expectation value of the $(0,0)$ and $(j,j)$ components of the energy-momentum tensor of a  field on   $|0(\theta, z)\rangle $,
\bea\label{energia}
\rho (z) &=& g_{00}   \langle 0(\theta, z)|: T^{00} (x):|0(\theta, z)\rangle\,,
\\\label{pressione}
p (z) &=& g_{jj}  \langle 0(\theta, z)|: T^{jj} (x):|0(\theta, z)\rangle\,.
\eea

Here $:...:$ denotes the normal ordering with respect to  $|0\rangle$ and no summation on the index $j$ is intended.

\section{Energy density of thermal vacuum and CMB temperature}

In the photon fields case, the explicit expressions of the energy momentum tensor density  $T^{\mu \nu}_{\ga}$ is
$T^{\mu\nu}_{\ga}  = - F^{\mu\alpha}F^{\nu}_{\alpha} + \frac{1}{4} g^{\mu\nu} F_{\alpha\beta}F^{\alpha\beta}$~\cite{Schweber,Leite}. As usual $F^{\al \beta} = \partial^{\beta} A^{\al} - \partial^{\al} A^{\beta}$, ($g^{\mu \nu} = (1, -1, -1,-1)$, $\mu = 0,1,2,3 \,$; $\hbar =1 = c$ will be used throughout the paper). The thermal vacuum condensate energy density is then
 \bea
\rho_{\gamma}(z) =
\int d^3 \; k \; {\Omega}_k \;  \langle 0(\theta, z)|:  a^\dag_k a_k :|0(\theta, z)\rangle\,,
\eea
where $\Omega_k = k$  for photons.
The result we obtain is
\bea\label{dark}
\rho_{\gamma}(z) =\,\frac{\pi^{2}\,k^{4}_{B}\,(1+z)^{4} T_{\gamma}^{4} }{15\, \hbar^{3}\,c^{3}}\,.
\eea

In a similar way, the contribution given to the  pressure by the thermal vacuum condensate of photons field is
%obtained by the expectation value of the $(j,j)$ component of the energy momentum tensor for photon fields $T^{jj}_{\ga}$ (no summation on the index $j$ is intended) on  $|0(\theta, z)\rangle$

\bea\label{pressure}
p_{\gamma}(z) = \frac{\pi^{2}\,k^{4}_{B}\,(1+z)^{4} T_{\gamma}^{4} }{45\, \hbar^{3}\,c^{3}}\, .
\eea
The equation of state is then $w_{\gamma}(z) = p_{\gamma}(z)/\rho_{\gamma}(z) = 1/3$, which is the equation of state of the radiation.
  Eqs.(\ref{dark}) and (\ref{pressure}) reproduce of course the results obtained by solving the Boltzmann equation for the  distribution function of photons in thermal equilibrium \cite{Kolb-Turner}.
  The advantage of the present computation is that the role of the boson condensate in obtaining such a result is underlined.
Taking the present CMB temperature, $T_{\gamma} = 2.72548 \pm 0.00057$ K,  and the present red shift of the universe, $z=0$, one obtains  the  value of the thermal vacuum energy density, $\rho_{\gamma} = 2 \times 10^{-51}GeV^{4} $, which of course coincides with the energy density of the CMB  \cite{Kolb-Turner}.

Leaving apart the photon case, we consider now massive boson and fermion fields. The energy momentunm tensor density is given by $T_{B}^{\mu\nu}(x) = \partial_{\mu}\phi(x) \partial^{\mu}\phi(x) -\frac{1}{2}  g_{\mu\nu} (\partial^{\rho}\phi(x) \partial_{\rho}\phi(x) - m^{2}\phi(x)^{2})\,
$ for  free real scalar fields $\phi$, and $T^{\mu \nu} = \frac{i}{2} \bar{\psi}\gamma^{\mu}\overleftrightarrow{\partial}^{\nu} \psi $ for free Majorana spinor fields $\psi$.

 At any epoch, the thermal vacuum energy and thermal pressure are given by Eqs.(\ref{energia}) and (\ref{pressione}), which in the case of  the field $\phi$ give
\begin{widetext}

\bea \label{T00Bos}
\rho_{B} & = &  \frac{1}{2}  \langle 0 ( \theta, z)|   : \Big[\pi^{2}( x) + \lf[\vec{\nabla} \phi( x)  \ri]^{2}
+ m^{2} \phi^{2}( x) \Big]: |0( \theta, z) \rangle\,;
\\ \label{TjjBos}
p_{B} & = &    \langle 0 ( \theta, z)| :\Big(\lf[ \partial_{j} \phi( x) \ri]^{2}+ \frac{1}{2}\Big[\pi^{2}( x)
 -   \lf[\vec{\nabla} \phi( x)  \ri]^{2}
 - m^{2}
\phi^{2}( x)  \Big] \Big): |0 (\theta, z) \rangle\,.
 \eea
\end{widetext}
In the case of the isotropy of the momenta $k_1 = k_2 =k_3$, these can be written as
\bea\label{enBos1}
 \rho_{B} & = &  \int \frac{d^{3} {\bf k}}{(2 \pi)^{3}}\, \Omega_{k }\, \langle 0 ( \theta, z)|   a^{\dagger}_{\bf k}\, a_{\bf k}\, |0( \theta, z) \rangle\,;
\\ \label{PreBos}\non
p_{B} & = &   \frac{d^{3} {\bf k}}{(2 \pi)^{3}}\,  \Big[\frac{1}{3} \frac{k^2}{\Omega_k}  \langle 0 ( \theta, z)|    a^{\dagger}_{\bf k}\, a_{\bf k}\,|0 ( \theta, z) \rangle
\\\non
& - & \lf( \frac{1}{3}  \frac{k^2}{\Omega_k} + \frac{1}{2} \frac{m^2}{\Omega_k} \ri)
 \langle 0 ( \theta, z)|\Big( a_{\bf k}\, a_{-\bf k} e^{-2 i \Omega_k t}\,
\\
& + &
 a^{\dagger}_{\bf k}\, a^{\dagger}_{ -\bf k} e^{2 i \Omega_k t}\Big)   |0 ( \theta, z) \rangle \Big]\,.
 \eea
Explicitly they become
\bea\label{energy-Bos}
\rho_{B} (z) & = & \frac{1}{2 \pi^{2}}  \int_{0}^{\infty} dk k^{2}\frac{\Omega_{k } }{exp\lf(\frac{\Omega_{k }}{k_{B}T_{\gamma}(1+z)}\ri) - 1}\,,
\\\label{pressure-Bos}
p_{B} (z)& = & \frac{1}{6 \pi^{2}}   \int_{0}^{\infty} dk k^{2}\, \Big[  \frac{k^2}{\Omega_k}\frac{1 }{exp\lf(\frac{\Omega_{k }}{k_{B}T_{\gamma}(1+z)}\ri) - 1}
\\\non
& - &  \lf( \frac{  k^2}{ \Omega_k} + \frac{3 m^2}{2 \Omega_k} \ri) \frac{exp\lf(\frac{\Omega_{k }}{2 k_{B}T_{\gamma}(1+z)}\ri) }{exp\lf(\frac{\Omega_{k }}{k_{B}T_{\gamma}(1+z)}\ri) - 1} \cos (2 \Omega_k t) \Big] \,.
 \eea

Notice that the vacuum energy density at thermal equilibrium  $\rho_{B} (z)$,  Eqs.(\ref{energy-Bos}), coincides with  the result obtained  by solving the Boltzmann equation for the particle Bose distribution function \cite{Kolb-Turner}.
The difference to the pressure $p_{B} (z)$  between the contributions coming from the vacuum condensate and the  ones coming
solely from the Bose distribution function appears in Eq.(\ref{pressure-Bos}).
The second term on the R.H.S. of Eq.(\ref{pressure-Bos}) appears due to the condensate of the physical vacuum contributing with non-vanishing values of
$\langle 0 ( \theta, z)|  a_{\bf k}\, a_{-\bf k}   |0 ( \theta, z) \rangle$ and
$
\langle 0 ( \theta, z)|
 a^{\dagger}_{\bf k}\, a^{\dagger}_{ -\bf k}  |0 ( \theta, z) \rangle$. Would the vacuum be the trivial one $|0\rangle$, these contributions would be identically zero.

By considering the present epoch, $z= 0$, $T =T_{\gamma}$, and by solving numerically the  integral in Eq.(\ref{energy-Bos}), one has the contribution to the vacuum energy given by $\rho_{B} \simeq 9 \times  10^{-52}GeV^4$ for masses less or equal than the CMB temperature $m \leq T_{\gamma}$, i.e. $m \leq 2.3 \times 10^{-4} eV$ (for example, possible candidates are axion-like   with
$m_{a} \in (10^{-3}- 10^{-6}) eV$). The maximum value of $\rho_{B}$ is obtained for   $m \ll  10^{-4} eV$. In this case, one has $\rho_{B} \simeq   10^{-51}GeV^4$.
Negligible values of $\rho_{B}$ are obtained for boson masses $m \gg 10^{-3} eV$.

%Therefore, apart from the contribution given by hypothetical particles as axion-like ones with estimated  masses
%$m_{a} \in (10^{-3}- 10^{-6}) eV$, the thermal vacuum contribution of bosons to the energy of the universe is completely negligible with respect to $\rho_{\gamma}$ and then to the estimated critical density of the present universe $\rho_{cr} \simeq 4.5 \times 10^{-47}GeV^{4}$.

 In the fermion case, the Fermi-Dirac distribution function is   obtained
\bea
\hspace{-4mm}{\cal N}^{F}_{a_{\bf k}}(\theta) = \sin^{2} \theta_k = \frac{1}{e^{\beta {\Omega}_k} + 1}\,.
 \eea
The thermal vacuum contribution to the  energy density and to the pressure, are
% \begin{widetext}

\bea \label{T00Fer}\non
\rho_{F} & = &  \frac{1}{2}   \langle 0 ( \theta, z)|   : \Big[-i \bar{\psi}\, \gamma_{j} \partial^{j}\, \psi + m \bar{\psi} \psi \Big]: |0( \theta, z) \rangle\,;
\\
\\ \label{TjjFer}
p_{F} & = &    \langle 0 ( \theta, z)| :\Big(  \frac{i}{2} \bar{\psi} \,\gamma_{j} \overleftrightarrow{\partial_{j}} \psi  \Big): |0 (\theta, z) \rangle\,,
 \eea
 %\end{widetext}
 respectively. In Eq.(\ref{T00Fer}),  the relation $\frac{i}{2} \bar{\psi} \,\gamma_{0} \overleftrightarrow{\partial_{0}} \psi = i \bar{\psi} \,\gamma_{0}  {\partial_{0}} \psi= -i \bar{\psi}\, \gamma_{j} \partial^{j}\, \psi + m \bar{\psi} \psi$ is used.
For Majorana fields,   Eqs.(\ref{T00Fer}) and  (\ref{TjjFer}) give
\bea \label{T00Fer1}
\rho_{F} & = &   \sum_{r} \int \frac{d^{3} {\bf k}}{2 \pi^{3}}\, \Omega_{k} \,\langle 0 ( \theta, z)|   \alpha^{r \dag}_{k}  \alpha^{r  }_{k}|0( \theta, z) \rangle\,;
\\ \label{TjjFer1}
p_{F} & = &   \frac{1}{3}\sum_{r} \int \frac{d^{3} {\bf k}}{2 \pi^{3}}  \frac{k^{2}}{\Omega_{k}}  \langle 0 ( \theta, z)|   \alpha^{r \dag}_{k}  \alpha^{r  }_{k}|0( \theta, z) \rangle\,,
 \eea
where $ \alpha^{r}_{k}$, $r =1,2$, is the annihilator of fermion field.
%The state equation of the fermion vacuum condensate is then,
 %  $w_{F} = 1/3$, for negligible mass term and isotropy of the momenta, and $w_{F} = 0$, for negligible gradient term.

 The explicit expressions of the energy density and pressure  are
\bea\label{energy-Fer}
\rho_{F} (z) &= & \frac{1}{\pi^{2} } \int_{0}^{\infty} dk k^{2}\frac{\Omega_{k } }{exp\lf(\frac{\Omega_{k }}{k_{B}T_{\gamma}(1+z)}\ri) + 1}\,,
 \\\label{pressure-Fer}
 p_{F}(z)  &= & \frac{1}{3 \pi^{2} } \int_{0}^{\infty} dk \frac{k^{4}}{\Omega_{k }}\frac{1 }{exp\lf(\frac{\Omega_{k }}{k_{B}T_{\gamma}(1+z)}\ri)+1}\,,
 \eea
respectively. These equations coincide with the energy density and pressure  obtained by solving the Boltzmann equation for the fermion distribution function \cite{Kolb-Turner}.
For $z = 0$ and masses $m \leq T_{\gamma}$, we find at $T= T_\gamma$ the maximum value of $\rho_{F}$, i.e. $\rho_{F} \sim 1.6 \times 10^{-51}GeV^{4}$ which is of the same order of CMB energy. The state equation is $w_{F} \sim 1/3$.
Condensates  of heavier fermions give negligible contributions to the universe energy. Only  particles with masses less or equal to $10^{-4} eV$, e.g. neutrinos, may give relevant contributions.

Taking into account such results, from  Eqs.(\ref{energy-Fer}) and (\ref{pressure-Fer})
 we compute the energy density and pressure for the three neutrino fields   at the   cosmic neutrino background (CNB) temperature  $T_{\nu} = 1.95 K$.
\footnote{The relic neutrino temperature $T_{\nu}$ is related to the one of CMB $T_{\gamma}$ by the relation \cite{Kolb-Turner}
\bea\non
\lf(\frac{T_{\nu}}{T_{\gamma}} \ri) = \lf(\frac{4}{11} \ri)^{1/2}\,.
\eea
This implies that since at the present epoch $T_{\gamma} = 2.725 K$, one obtains $T_{\nu} = 1.95 K$.}

 For $z=0$ and neutrino masses  $m_i \sim 10^{-4} eV$,  the maximum value of the energy density turns out to be  $\rho_{\nu} \sim  0.5 \times  10^{-51}GeV^{4}$, with state equation   $w_{\nu} \sim 1/3$.
Larger  neutrinos masses would give
 negligible contributions to  $\rho_{\nu}$.
 Adopting as customary  $ \rho_{\nu} \leq \rho_{\gamma} $, and taking the    mass  $m_{\nu,1} \sim 10^{-4}eV$ (which leads to $ \rho_{\nu} \leq \rho_{\gamma} $) to by the lighter neutrino mass, one can derive $m_{\nu,2}$ and $m_{\nu,3}$ from the hierarchical neutrino model and $\Delta m_{12}^{2} = 8 \times 10^{-5}eV^2$ and $\Delta m_{23}^{2} = 2.7 \times 10^{-3}eV^2$. The result is
 $m_{\nu,2} = 9  \times 10^{-3}eV$  and $m_{\nu,3} = 5.3 \times 10^{-2}eV$, and thus $\sum m_{\nu} = 6 \times 10^{-2}eV$,  as it should be  in agreement with its estimated lower bound.

%If one assumes that the contribution of the neutrino thermal vacuum is bounded from above by the photon thermal vacuum contribution, i.e. $ \rho_{\nu} \leq \rho_{\gamma} $, and that the above mass value  represents the mass of the lighter neutrino mass neutrino, $m_{\nu,1} \sim 10^{-4}eV$, considering the hierarchical neutrino model, according to which $\Delta m_{12}^{2} = 8 \times 10^{-5}eV^2$ and $\Delta m_{23}^{2} = 2.7 \times 10^{-3}eV^2$,
%one can derive the masses $m_{\nu,2}$ and $m_{\nu,3}$  of the other neutrinos.
%One has, $m_{\nu,2} = 9  \times 10^{-3}eV$  and $m_{\nu,3} = 5.3 \times 10^{-2}eV$, then $\sum m_{\nu} = 6 \times 10^{-2}eV$, which is in agreement with the estimated lower bound on the sum of the three neutrino masses.

 \section{Fractal  structure of the thermal states}

Finally, we show that the thermal vacuum $|0(\theta)\rangle$ has a fractal self-similar structure.
Let us consider the time dependent case $\theta = \theta (t)$.
%temperature, $\beta = \beta(t) = 1/(k_{B}T(t))$.
We will use the notation $|0(\theta(t))\rangle \equiv |0 (t) \rangle$. The boson vacuum $|0(t)\rangle$ provides the quantum representation of the system of couples of damped/amplified oscillators  \cite{Celeghini:1992yv}
\bea \lab{eqxy1}
m \ddot x + \gamma \dot x + k x  &=& 0 ,\\
\lab{eqxy2}
m \ddot y - \gamma \dot y + k y  &=& 0 ,\\
\lab{Lxy}
 L = m\dot x \dot y + {\gamma\over 2} (x \dot y -\dot x y) &-& k x\,y ,
\eea
where ``dot'' denotes time derivative, $m$, $\ga$ and $\kappa$ are positive real constants
 and $L$ is the Lagrangian from which Eqs. (\ref{eqxy1}) and (\ref{eqxy2}) are derived.

To see indeed how $|0(t)\rangle$ is obtained, one proceeds to the canonical quantization of the system described by Eqs. (\ref{eqxy1}) - (\ref{Lxy}) and assumes that the canonical commutation relations hold
$[\, x , p_{x} \, ] = i\, \hbar = [\, y , p_{y} \,]~,~   [\, x ,
y \,] = 0 = [\, p_{x} , p_{y} \, ]$.
The corresponding sets of annihition and creation operators are
\bea
\lab{aop}
\alpha  &\equiv& \left ({1\over{2 \hbar \Omega}} \right )^{1\over{2}} \left (
{{p_{x}}\over{\sqrt{m}}} - i \sqrt{m} \Omega x \right ); \\
%a^{\dagger} \equiv \left ({1\over{2 \hbar \Omega}} \right )^{1\over{2}}
%\left ({{p_{x}}\over{\sqrt{m}}} + i \sqrt{m} \Omega x \right )
%\ee
%\be
\lab{bop}
\widetilde{\alpha} &\equiv& \left ({1\over{2 \hbar \Omega}} \right )^{1\over{2}} \left (
{{p_{y}}\over{\sqrt{m}}} - i \sqrt{m} \Omega y \right ); ~
%%
%b^{\dagger} \equiv \left ({1\over{2 \hbar \Omega}} \right )^{1\over{2}}
%\left ( {{p_{y}}\over{\sqrt{m}}} + i \sqrt{m} \Omega y \right )
\eea
 with $[\, \alpha , \alpha^{\dagger} \,] = 1 = [\, \widetilde{\alpha} , \widetilde{\alpha}^{\dagger} \,]$, $\quad
[ \,\alpha , \widetilde{\alpha} \,] = 0 = [\, \alpha , \widetilde{\alpha}^{\dagger} \,]$.
 The canonical linear transformations $ a  \equiv
(1/{\sqrt 2}) ( \alpha + \widetilde{\alpha}  )$, $b  \equiv (1/{\sqrt 2})
( \alpha  - \widetilde{\alpha} )$ are introduced.
It is found~\cite{Celeghini:1992yv} that the time evolution of the system ground state (the vacuum) leads out of the Hilbert space of the states, and thus the proper quantization setting is the one of the quantum field theory (QFT). One has therefore to consider operators $a_{\bf k}$, $b_{\bf k}$ and their hermitian conjugates, so to perform, as customary in QFT,
 the continuum momentum limit (or the infinite volume limit) by use of the relation $ \sum_{\kappa} \rar
(V/{(2 \pi)^{3}}) \int \! d^{3}{\kappa}$ at the end of the computations.
 The Hamiltonian $H$ of the system is found to be~\cite{Celeghini:1992yv}
$H =  H_{0} +  H_{I}$, with
\bea
\label{HAB2}
H_{0} = \sum_{\bf k}\hbar \Omega_{k} ( a^{\dagger}_{\bf k} a_{\bf k} - b^{\dagger}_{\bf k} b_{\bf k})\,,
\\
H_{I} = i \sum_{\bf k}\hbar \Gamma_{k} ( a^{\dagger}_{\bf k} b^{\dagger}_{\bf k} - a_{\bf k} b_{\bf k})\,,
\eea
where it has been used $\theta_{\kappa} (t) = \Ga_{k}\, t \equiv ({{\ga_{k}}/{2 m}})\, t $ for each $\kappa$-mode. The group structure is the one of the $SU(1,1)$, $ [\, H_{0}, H_{I}\, ] = 0 $ and the Casimir operator ${\cal C}$ is given by
${\cal C}^{2} = (1/{4}) ( a^{\dagger}_{\bf k} a_{\bf k} - b^{\dagger}_{\bf k}b_{\bf k})^{2}$. The initial condition of positiveness for the eigenvalues of $H_{0}$ are thus protected against transitions to negative energy states. One then finds that the time evolution of the vacuum $|0\rangle$ for $a_{\bf k}$ and $b_{\bf k}$  is controlled by $H_{I}$ and given by $|0(\theta(t)) \rangle  = e^{  - i t {H \over{\hbar}} } |0\rangle = e^{  - i t {H_{I} \over{\hbar}} } |0 \rangle$ which gives in fact Eq. (\ref{(2.12)}).

One also finds that $| 0(t)\rangle $ turns out to be a squeezed coherent state characterized by the $q$-deformation of Lie-Hopf algebra  and provides a representation of the CCR at finite temperature which is equivalent~\cite{Celeghini:1992yv} to the Thermo Field Dynamics representation $\{ |0(\beta)
\rangle \}$~\cite{Takahashi:1974zn,Umezawa}. In the limit of quasi-stationary case with $\beta (t)$ slowly changing in time, minimization of the free energy gives again the
Bose-Einstein distribution function  Eq. (\ref{(2.13a)}).

Indeed, let us now introduce the functional ${\cal F}_{a}$ for the $a-$modes
\bea\label{F}
{\cal F}_{a} \equiv \langle 0(t)
|\left(  H_{a} - \frac{1}{\beta} S_{a} \right)| 0(t)\rangle\,,
\eea
where $ H_{a} $  is the free Hamiltonian relative to the   $a-$modes,
$H_{a} = \sum_{\bf k} \hbar \Omega_{ k}\, {a}_{\bf k}^{\dagger} a_{\bf k} $, and $S_{a}$ is given by
\bea\label{S}
S_{a} \equiv - \sum_{\bf k}\left\lbrace  {a}_{\bf k}^{\dagger} a_{\bf k} \ln \sinh^{2}(\theta) - a_{\bf k} {a}_{\bf k}^{\dagger}\ln \cosh^{2}(\theta) \right\rbrace \,.
\eea
 Inspection of Eqs.(\ref{F}) and (\ref{S}) suggests that ${\cal F}_{a}$
and $S_{a} $ can be considered as free energy and the entropy, respectively.
Minimization of the functional ${\cal F}_{a}$, $\frac{\partial {\cal F}_{a}}{\partial \theta_{k}(t)} = 0$, $\forall k$~\cite{Takahashi:1974zn,Umezawa} (we consider $\hbar =c =1$) then leads to Eq.(\ref{(2.13a)})
which is the Bose-Einstein distribution function for $a_{\bf k}$.  The first principle of thermodynamics at constant temperature  can be then expressed as
\bea
d {\cal F}_{a} = d {\cal E}_{a} - \frac{1}{\beta}{\cal S}_{a} =0\,,
\eea
where, the change in time of the particle condensed in the vacuum turns out into heat dissipation $d Q = \frac{1}{\beta} d {\cal S} $
\bea
d {\cal E}_{a} = \sum_{\bf k} \hbar \, \Omega_{\bf k}\,  d {\dot{\cal N}}_{a}^{\bf k}(t)\, dt = \frac{1}{\beta} d {\cal S} =d Q   \,,
\eea
 where ${\dot{\cal N}}_{a}^{\bf k}(t)$ denotes the time derivative of ${\cal N}_{a}^{\bf k}(t)$.

We now remark that the system of  Eqs.~(\ref{eqxy1}) and (\ref{eqxy2})) posses self-similarity properties.
To see this, let us put
\bea
\frac{1}{2}[z_1 (t) + z^{*}_2 (-t)] &=& x(t)\\
\frac{1}{2}[z^{*}_1 (-t) + z_2 (t)] &=& y(t)
\eea
 with  $z_1 (t) = r_0 \,  \, e^{- \,i \, \Om \, t}\, e^{- \Ga t }$ and $z_2  (t) = r_0 \,  \, e^{+ \,i \, \Om\,t}\, e^{ + \Ga \, t }$,   $\Ga \equiv {\ga / 2 m}$ and $\Om^2 = (1/m) (\ka- \ga^2 /4m)$,  $\ka > \ga^2 /4m$. Then we see that
 Eqs.~(\ref{eqxy1}) and (\ref{eqxy2}) can be rewritten as~\cite{PLA2012}
\bea \lab{losp2.10} m \, \ddot{z}_1 \, + \, \ga \, \dot{z}_1 \, + \, \kappa \, z_1  &=& 0 ,\\ \lab{losp2.10b}
m \, \ddot{z}_2  \, - \, \ga \, \dot{z}_2 \, + \, \kappa \, z_2  &=& 0 .
\eea
Solutions of Eqs.~(\ref{losp2.10}) and (\ref{losp2.10b}) are in fact  $z_1 (t) = r_0 \,  \, e^{- \,i \, \Om \, t}\, e^{- \Ga t }$ and $z_2  (t) = r_0 \,  \, e^{+ \,i \, \Om\,t}\, e^{ + \Ga \, t }$  and they describe the parametric time evolution of clockwise  and the anti-clockwise logarithmic spirals, $r = r_{0} e^{- d\alpha}$ and $r = r_{0} e^{d\alpha}$, with $\alpha (t) = \Gamma \, t/d$ and $\Omega \, t = \Gamma \, t/d = \alpha (t)$~\cite{PLA2012}.
%with $\theta(t) =   \ga \, t/(2 \, m \, d)  \equiv  \Ga \, t/d $
%up to an arbitrary additive constant;

Thus, Eqs.~(\ref{eqxy1}) and (\ref{eqxy2}) (or equivalently Eqs.~(\ref{losp2.10}) and (\ref{losp2.10b})), whose quantum representation is provided by $| 0(t)\rangle $, are found to describe the self-similar fractal structure of their logarithmic spiral solutions~\cite{Peitgen,Andronov}. This establish the link  between the $SU(1,1)$ coherent states and fractal-like self-similarity~\cite{PLA2012}.
The relation of the photon energy-momentum tensor  $T^{\mu\nu}_{\ga}$ with Eqs.~(\ref{eqxy1}) and (\ref{eqxy2}) can also be shown. For details see~\cite{PLA2012}. Similar discussions can be done for the fermion vacuum.

\section{Conclusions}

We have studied  the thermal vacuum structure at the temperature of the CMB. In the  framework of TFD, the the energy momentum tensor density of photon has expectation value on the vacuum  which agrees with the energy density and pressure of the CMB.
In the  case of neutrinos and thermal states at the temperature of the CNB consistency has been verified with    the estimated lower bound of the sum of the active neutrino masses.
The fractal self-similar structure of the thermal   vacuum has been also discussed.

\section{Conflict of Interests}

The authors declare that there is no conflict of interests
regarding the publication of this paper.

\section{Acknowledgements}
Partial financial support from MIUR is acknowledged.

 \end{document}